\documentclass[lettersize,journal]{IEEEtran}
\usepackage{amsmath,amsfonts}
\usepackage{algorithmic}
\usepackage{algorithm}
\usepackage{array}
\usepackage[caption=false,font=normalsize,labelfont=sf,textfont=sf]{subfig}
\usepackage{textcomp}
\usepackage{stfloats}
\usepackage{url}
\usepackage{verbatim}
\usepackage{graphicx}
\usepackage{multirow}
\usepackage{cite}
\usepackage{lineno}
\usepackage{float}
\usepackage{xcolor}
\definecolor{postcode}{named}{black}
\begin{document}
\title{Sensing and Understanding the World over Air: \\ A Large Multimodal Model for Mobile Networks}
\author{Zhuoran Duan\textsuperscript{*},
        Yuhao Wei\textsuperscript{*},
        Guoshun Nan~\IEEEmembership{Member,~IEEE,}
        Zijun Wang,
        Yan Yan,
        Lihua Xiong,
        Yuhan Ran,
        Ji Zhang,
        Jian Li,
        Qimei Cui~\IEEEmembership{Senior Member,~IEEE,}
        Xiaofeng Tao~\IEEEmembership{Senior Member,~IEEE,}
        \\
        Tony Q.S. Quek~\IEEEmembership{Fellow,~IEEE}

\thanks{

Zhuoran Duan, Guoshun Nan, Zijun Wang, Yan Yan, Lihua Xiong, Jian Li, Qimei Cui, and Xiaofeng Tao are with the National Engineering Research Center for Mobile Network Technologies, Beijing University of Posts and Telecommunications (BUPT), Beijing 100876, China, and also with Beiyou Shenzhen Institute, Shenzhen, China. (\textit{corresponding author: Guoshun Nan.})

Yuhao Wei is with the School of Cyber Security, University of Chinese Academy of Sciences (UCAS){\color{postcode}}.

Yuhan Ran is with the Electrical and Electronic Engineering School at the University of Bristol.

Ji Zhang is with the China Telecom Co., Ltd.

Tony Q.S. Quek is the Cheng Tsang Man Chair Professor with the Singapore University of Technology and Design (SUTD).

\textsuperscript{*}These authors contributed equally to this work.
}
}
\maketitle

\begin{abstract}
Large models (LMs), such as ChatGPT, have made a significant impact across diverse domains and hold great potential to facilitate the evolution of network intelligence. Wireless-native multi-modal large models (WMLMs) can sense and understand the physical world through multi-modal data, serving as a key enabler that integrates communication, sensing, and intelligence, and thus they can boost various smart services to billions of users. However, research on WMLMs remains in its infancy, and the construction of domain-specific multi-modal large models for wireless networks is still underexplored. In this paper, we outlines the key characteristics of WMLMs and summarizes existing methods, on the basis of which a wireless-native multimodal training paradigm is proposed. Specifically, we constructed a GPT-style WMLM model and trained it on a real-world large-scale dataset, leveraging wireless signals as an anchor modality for contrastive learning. Our approach demonstrates outstanding performance compared with existing small-scale models and large multi-modal models, validating the feasibility of using wireless signals as a universal modality and highlighting WMLM’s potential to emerge as a new paradigm for future wireless networks.

\end{abstract}


\section{Introduction}

\label{sec:1}
The advent of large AI models (LMs) such as ChatGPT has propelled network intelligence into a new evolutionary phase. These remarkable enablers are poised to revolutionize future wireless networks through their advanced performance and generalization capability. Recent research~\cite{LLM4Beam,LLM4CP,CSI_BERT} has demonstrated that LMs achieve outstanding performance in wireless communication tasks (e.g., beam prediction and channel estimation). Meanwhile, 6G emphasizes the integration of communication, sensing, and intelligence~\cite{IMT-2030}. The massive sensing and communication data will endow models with machine synesthesia~\cite{multi-modal sensing and comm}, enabling them to serve applications across various scenarios. Therefore, Wireless native multi-modal large models (WMLMs) hold the promise of bridging communication, sensing and intelligence, paving the way for advanced information processing for future networks. 

\begin{figure*}[t!]
  \centering
  \includegraphics[width=1.1\textwidth,trim={10 0 40 0}, clip]{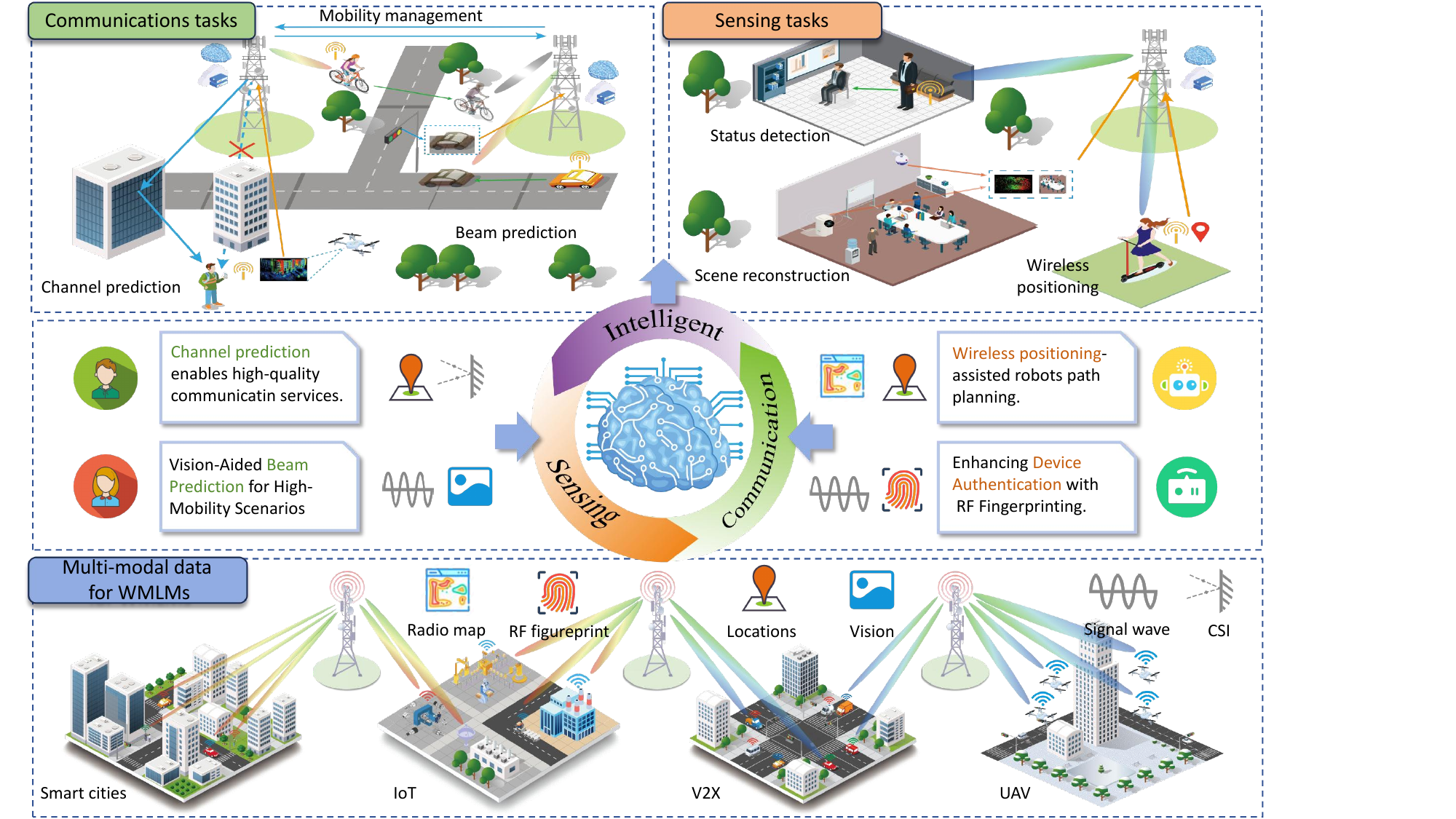}

  \caption{WMLMs for network. WMLMs integrate sensing, communication, and intelligence, providing networks with versatile intelligent information processing capabilities. WMLMs correlate patterns in multi-modal sensing and communication data, thereby empowering various downstream tasks in multiple scenarios.}
  \label{fig1}
\end{figure*}

WMLMs are expected to achieve perception of the physical world through massive wireless sensors, which enables them to handle multiple network-native tasks. In summary, Fig.\ref{fig1} illustrates how WMLMs enhance future wireless networks. Taking advantage of the feature extraction and adjustment capability of artificial intelligence, different types of modalities (e.g., vision, channel state information, RF figuerprint) have been widely used for tasks which are not easily performed. Changes in the physical world (e.g., channel state variations and object motion trajectories) can be characterized by multi-modal data features. The synergy between these modalities has given rise to various applications, such as vision-aided beam prediction and fingerprint-based localization. WMLMs can further enhance the capability of data fusion. Moreover, this paradigm involves constructing foundation models through pre-training on vast datasets, followed by rapid adaptation to various downstream tasks via fine-tuning, few-shot learning, and zero-shot learning methods. Compared to small-scale models in wireless networks, large-scale models exhibit robust generalization and advanced intelligence, significantly enhancing performance across diverse tasks in multiple scenarios.


Current research on wireless native large scale models remains in the exploratory phase, with existing studies~\cite{LLM4Beam,LLM4CP,CSI_BERT,GenAI4PHY} discussing their feasibility from multiple perspectives including applications, deployment, and training. These works outline a viable pathway toward achieving versatile intelligent information processing. However, there is a lack of detailed discussion on MLMs that enable communication and sensing, particularly regarding how these models can effectively integrate heterogeneous multi-modal data from wireless networks and enhance both communication and sensing tasks. This paper investigates and discusses three key aspects of WMLMs to address these challenges:

\textbf{The characteristics of WMLMS:} The design of WMLMs is constrained by the characteristics of wireless networks, particularly their stringent latency and limited resources.

\textbf{Multi-modal fusion:} As multi-modal data forms the basis for environmental perception, developing effective fusion mechanisms is critical, though it remains in its early stages.

\textbf{Representations of downstream tasks:} WMLMs should demonstrate both task-processing versatility and generalization capability across diverse network-native applications.

Based on the above discussions, we propose a two-stage training framework to validate the feasibility of Wireless Multi-modal Large Models (WMLMs), comprising: (1) modality-specific alignment and (2) task-specific adaptation. This two-stage paradigm ensures effective model deployment and application in wireless network environments. We then carry out extensive experiments on real-world datasets to verify the feasibility of our approach. Finally, we outline future research directions, discuss potential challenges, and present concluding remarks. The main contributions of this paper are summarized as follows:

\begin{itemize}
    \item We present a comprehensive perspective on WMLMs for wireless networking, covering desired characteristics, multi-modal data fusion, and multi-task representation learning. This lays out a concrete and feasible pathway toward realizing WMLMs.
    \item Inspired by these guidelines, we propose a two-stage training and deployment paradigm for WMLMs applicable across diverse network scenarios.
    \item We conduct extensive experiments to evaluate the performance of our WMLMs on a real-world ISAC dataset, and valuate the feasibility of using wireless signals as a universal modality for future networks.
\end{itemize}

\section{Key characteristics of WMLMs}
\label{sec:2}

\begin{figure}[t!]
  \centering
  \includegraphics[width=1\textwidth,trim={10 160 0 0}, clip]{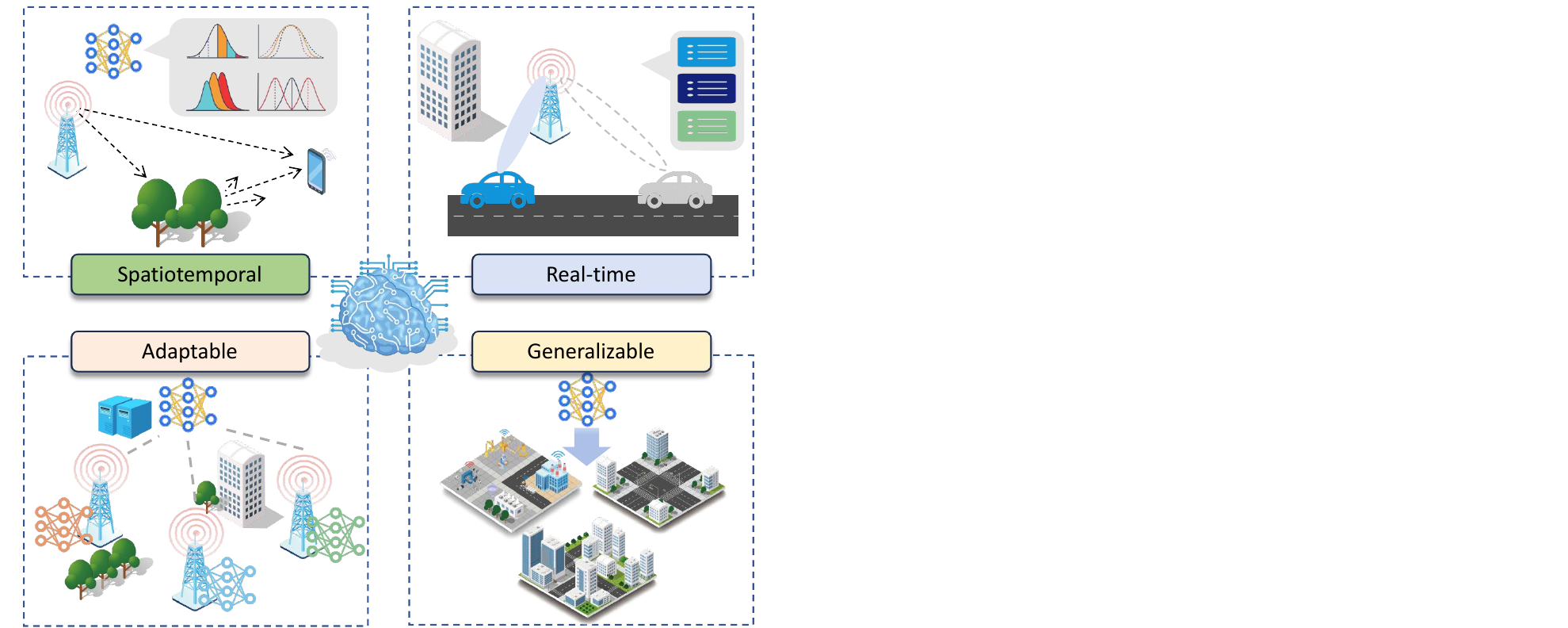}

  \caption{Key characteristics of WMLMs in future networking: spatiotemporality, real-time performance, adaptability, and generalization}
  \label{fig2}
\end{figure}

While existing MLMs have achieved notable success across various domains, their deployment in wireless networks imposes more stringent demands, particularly in terms of robustness, operational efficiency, and adaptability to dynamic environments. To cope with the inherent challenges of wireless systems, WMLMs must undergo adaptations beyond conventional MLM paradigms. As illustrated in Fig.~\ref{fig2}, we summarize four key characteristics of WMLMs:

\subsection{Spatiotemporal}

In wireless networks, multi-source data (e.g., spatial locations, channel state information, etc.) can characterize the physical state variations in the real world, where complex spatio-temporal-frequency coupling relationships typically exist among these modalities. Consequently, WMLMs must possess two core capabilities: (1) accurately capturing long-term dependency characteristics of complex patterns in time series, and (2) effectively extracting physically meaningful spatial features. Recent studies~\cite{LLM4CP,CSI_BERT} have shown that text-pretrained foundation models can effectively enhance downstream tasks in wireless communication. These studies demonstrate that large scale transformer-based architectures can effectively capture temporal-spatial dependencies in sequential data and generalize to communication tasks such as channel prediction. Meanwhile, pretraining foundation models with spatiotemporal data is a promising paradigm that avoids the model's reliance on text and better aligns with real-world network conditions. This paradigm can fully leverage the vast amounts of communication and sensing data in wireless networks for training, establishing spatiotemporal relationships among multi-modal data. Some current studies~\cite{timer} have already demonstrated the feasibility of training general-purpose large-scale time series models. However, research specifically targeting wireless networks remains unexplored.

\subsection{Real-time}

With the continuous evolution of wireless networking, emerging services such as immersive communications impose increasingly stringent latency requirements on networks. There exists an inherent trade-off between real-time performance and the computational overhead of large-scale models. Particularly under the mainstream autoregressive generation paradigm, where processing time scales linearly with output length, thereby constraining the overall real-time capability of communication systems. Existing studies~\cite{LLM4CP} demonstrate that models at the scale of GPT-2 can achieve millisecond-level inference latency on general-purpose computing platforms. However, more advanced LLMs (with parameter counts exceeding 3B) still suffer from second-level inference delays. While techniques like model distillation, quantization, and AI-specific hardware acceleration can mitigate real-time challenges, pursuing maximally large and universal models may not be the optimal strategy for latency-sensitive wireless communication systems, especially given energy and resource constraints in wireless networks. A careful balance between model scale and performance is essential. Thus, at the current stage, WMLMs are more likely to serve as domain-specific "mid-scale models" to enhance network performance.

\subsection{Adaptable}

In the real world, factors such as complex scattering environments and non-ideal propagation conditions, including weather variations and nonlinear hardware losses, introduce heterogeneity among different wireless access nodes. Shifts in statistical characteristics may degrade model performance in downstream tasks such as environmental sensing and communication. Wireless machine learning models must possess rapid scene adaptation capabilities to address challenges arising from environmental dynamics. Wireless edge networks offer promising solutions for these models, including edge device fine tuning, federated learning, and continual learning, which can enhance model performance across diverse wireless nodes. However, these approaches impose higher computational demands on edge networks. Furthermore, with the expansion into higher frequency bands such as Thz and visible light, there is an urgent need to explore the correlations between multi-modal data and their impact on model performance.

\subsection{Generalizable}

Generalization capability is a key feature of WMLMs. The ability to generalize outstanding performance across multiple scenarios and tasks is its most significant advantage compared to small-scale models, and it is also the foundation for MLMs to serve as a backbone for universal information processing. Scaling laws has shown that as the scale of training data and model parameters increases, MLMs exhibit efficient performance on entirely new data, demonstrating zero-shot learning capabilities. Similarly, potential correlations exist among data from different scenarios and modalities in wireless networks, and this trend is expected to hold. With the ongoing evolution toward integrated space-air-ground networks, WMLMs are expected to achieve cross-scenario capability generalization, enabling unified information processing across terrestrial and non-terrestrial domains.

\begin{table*}[t]
\centering
\renewcommand{\arraystretch}{2} 
\caption{Potential multi-modal data fusion and multi-task representation methods for WMLMs}
\label{tab}
\begin{tabular}{|>{\centering\arraybackslash}m{3cm}|>{\raggedright\arraybackslash}m{5cm}|>{\raggedright\arraybackslash}m{9cm}|} 
\hline
\multicolumn{1}{|c|}{\textbf{Aspects}} & \multicolumn{1}{c|}{\textbf{Research Directions}} & \multicolumn{1}{c|}{\textbf{Features}} \\
\hline
\multirow{3}{*}{Multi-modal data fusion}
& LLM-based reprogram & Adapting pre-trained LLMs for downstream tasks involving cross-modal communication. \\
\cline{2-3}
& Modality alignment & Using the specified modality as an anchor to guide different modalities in learning a shared representation space, enables feature-level modality fusion. \\
\cline{2-3}
& Unified Multi-modal representation & Unified multi-modal training jointly learns from diverse modalities during pre-training, enabling wireless-native AI with deeper cross-modal understanding. \\
\hline
\multirow{3}{*}{Multi-task representation} 
& Classification & Jointly learns device identification, activity recognition, or anomaly detection by sharing spatiotemporal representations.  \\
\cline{2-3}
& Forecast & Forecasts channel states, user mobility, or network traffic via unified sequence modeling. \\
\cline{2-3}
& Generation & Autoregressive generation using multimodal sensory data enables diverse tasks such as 3D scene reconstruction and channel modeling. \\
\hline
\end{tabular}
\end{table*}

\section{Multi-modal Data Fusion and Multi-Task Representation of WMLMs}
\label{sec:3}

WMLMs require novel architectures and training paradigms, unlike small-scale and single-task-oriented AI models. In particular, current MLMs constructed following the Generative Pre-trained Transformer (GPT) paradigm are expected to serve as the foundation for WMLMs. These models incorporate entirely new paradigms such as pre-training, post-training, and fine-tuning. We categorize the construction methods of WMLMs into two aspects: multi-modal data fusion and multi-task representation, based on data perception and task processing, as summarized in Table~\ref{tab}, along with existing approaches.

\subsection{Multi-modal Data Fusion}

The current multimodal data fusion paradigm for WMLMS remains in the exploratory stage, with joint feature embedding of multimodal data being the most common fusion approach for MLMs. Currently, LMs process different modalities (e.g., text, images) by splitting them into patches and mapping them into corresponding feature spaces, providing transformer-based backbone networks with multimodal feature representations. The fusion paradigms applicable to WMLMS can be categorized into three types:

\textbf{LLM-based reprogramming}: Pre-trained LLMs, leveraging large-scale text data, exhibit superior long-sequence understanding and modeling capabilities. Existing studies~\cite{LLM4Beam,LLM4CP,CSI_BERT} explore LLM-based adaptation for wireless downstream tasks, which is a highly competitive approach. LLMs can generalize to wireless network tasks without additional training overhead by fine-tuning on limited supervised data. 

\textbf{Modality alignment}: Some studies ~\cite{ImageBind, Clip} adopt fixed modalities (e.g., images, text) as anchors, aligning them with other modalities via contrastive learning to establish a shared representation space. This method offers flexibility in adapting to network and scenario changes while enabling modality extension. By exploiting inter-modal correlations and complementarity, multi-perspective learning enhances WMLMs' perception of channels, users, and physical environments. Moreover, wireless communication data can serve as an alignment anchor, providing a native solution for multi-modal fusion. For instance, spatiotemporal-frequency channel information from uplink feedback can be combined with vision or 3D point clouds to improve channel variation modeling. Advances in sensing technology further strengthen modality association capabilities through richer environmental data.

\textbf{Unified multi-modal representation}: Unlike pre-training LMs first and then performing modality alignment, unified multimodal training simultaneously processes and learns from multiple modalities during pre-training. This approach simplifies the training pipeline while enhancing the model's deep understanding and alignment of multi-modal information. For wireless communication networks, this scheme is expected to build wireless-native WMLMs. A foundational model for wireless communication and sensing requires joint training on large-scale multi-modal data. With continuous advancements in computing, emerging technologies such as diffusion models and digital twins are anticipated to generate vast amounts of data for model pre-training.

\subsection{Multi-Task Representation}

The downstream tasks of WMLMs are highly diversified. Equipped with rich environmental perception capabilities, MLMs are not limited to the two fundamental tasks in scenarios—perception enhancement and communication enhancement—but can also extend to domains such as radio monitoring, anomaly detection, and more. Due to the spatiotemporal nature of models and communication downstream tasks, we outline three task types in spatiotemporal data processing (classification, forecast, generation), which cover the majority of current AI tasks for wireless networking. 

In specific domains, these tasks exhibit inherent correlations. For instance, CSI and visual data can be jointly used to estimate the state of the communication environment for adaptive adjustments, as well as for localization. Additionally, tasks demonstrate complementarity; for example, CSI prediction and imputation can enhance channel state estimation, leading to more accurate and rapid localization and communication. Recent studies represent~\cite{LLM4WM,NetLLM,Agent} different downstream tasks by constructing multi-task heads or Mixture of Experts (MoE) models, which is a highly promising task representation approach for WMLMs. Therefore, multi-task representation in wireless scenarios holds significant potential. MLMs can learn task representations through unsupervised, semi-supervised, or supervised learning. Furthermore, a multi-stage training paradigm can be adopted to build a universal foundational model, enhancing performance across various downstream tasks.

\begin{figure*}[t!]
  \centering
  \includegraphics[width=\textwidth,trim={10 0 90 0}, clip]{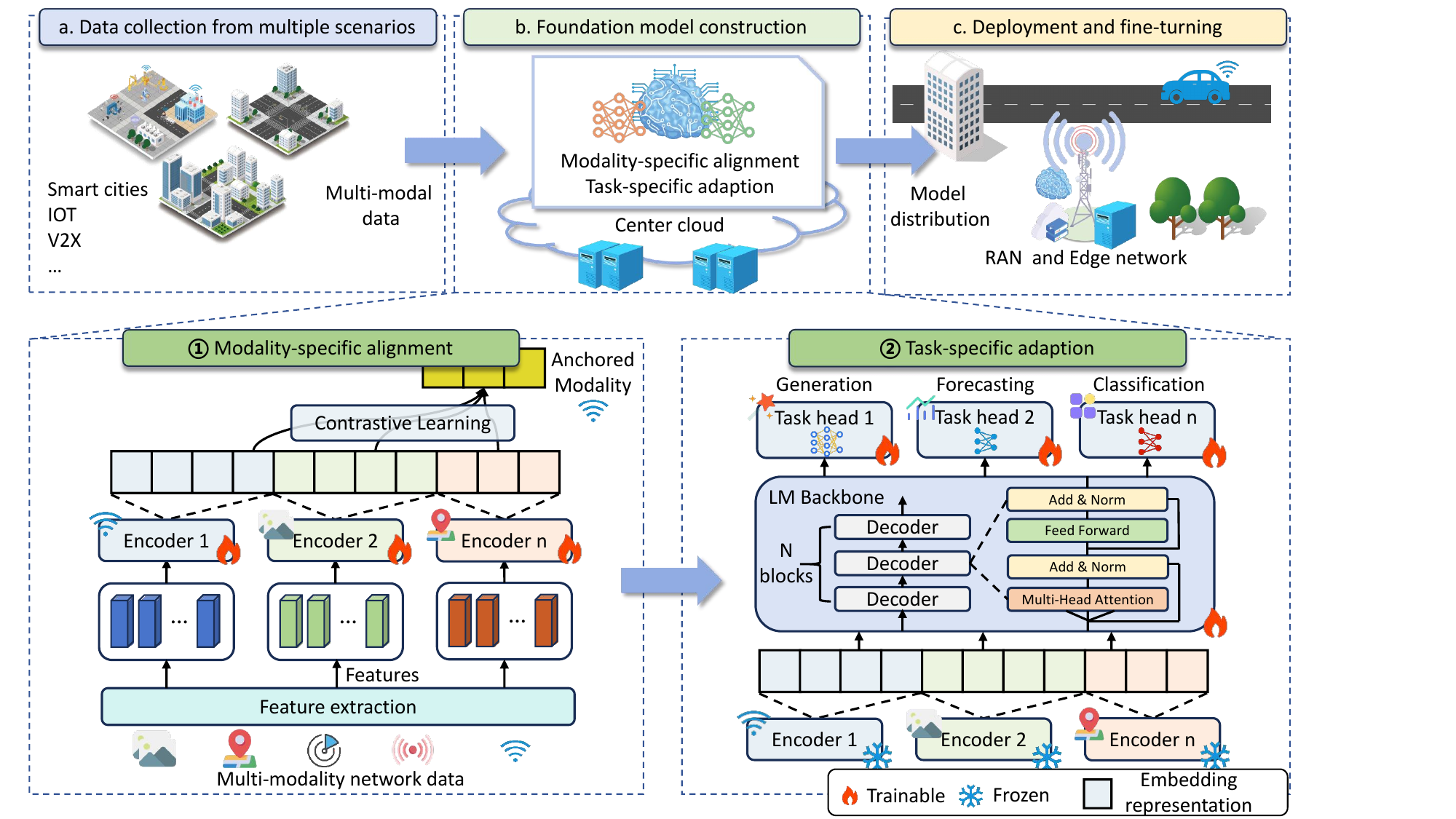}
  \caption{\textbf{Three-stage construction process for WMLMs:} a. Data acquisition from wireless communication networks, b. Cloud-based model construction, and c. Model deployment to Edge networks with fine-tuning. \textbf{Two-stage training paradigm for WMLMs:} \textcircled{1}. Cross-modal alignment via contrastive learning and \textcircled{2}. Downstream task adaptation.}
  \label{fig3}
\end{figure*}

\section{Two stage PARADIGM FOR WMLMS}
\label{sec:4}
According to the 6G vision, we assume a network comprising multiple scenarios and a center cloud. These scenarios include smart cities, industrial IoT, and vehicular networks. Within these scenarios, diverse communication and sensing devices collect massive amounts of multi-modal data, which can be transmitted to the center cloud. The cloud is equipped with extensive server clusters, supporting large-scale AI training. Edge networks are provisioned with computing platforms capable of deploying AI models and conducting small-scale training. Base stations and terminals are equipped with ISAC functionalities, and different scenarios feature base stations and terminals with various sensor devices.

In response to the aforementioned scenario, we proposed a universal training framework for WMLMs. This comprehensive paradigm encompasses three stages: multi-scenario data collection, foundational model construction, and model deployment and fine-tuning. Building upon this framework, we propose a two-stage training paradigm specifically designed for the construction of a general foundational model. The framework and paradigm described above collectively offer a viable solution for the training and deployment of WMLMs tailored for intelligent ISAC.

In intelligent wireless network scenarios, the network collects a vast array of communication and sensing data across different modalities. These data originate from various devices such as base stations, roadside detection units, and terminals, with sensor data encompassing modalities like RF-signals, images, positioning, and 3D point clouds. Data of diverse modalities are transmitted from the edge network to the central cloud. The central cloud trains foundational models tailored to specific scenarios based on environmental conditions, sensor types, and task requirements, leveraging the extensive computational resources of its large-scale computing clusters. Subsequently, these foundational models are distributed from the central cloud to various edge network nodes and deployed on the intelligent RAN (Radio Access Network) side for real-time inference. Since the disseminated models are customized for specific scenarios, they are well-suited for edge inference, obviating the need for additional training and deployment overhead, thereby ensuring real-time inference while maintaining general intelligence. Moreover, the intelligent RAN side possesses certain edge computing capabilities, enabling model fine-tuning and dynamic learning at the edge to adapt to the ever-changing physical environment.

\begin{figure*}[b!]
  \centering
  \includegraphics[width=\textwidth,trim={5 140pt 220 5pt}, clip]{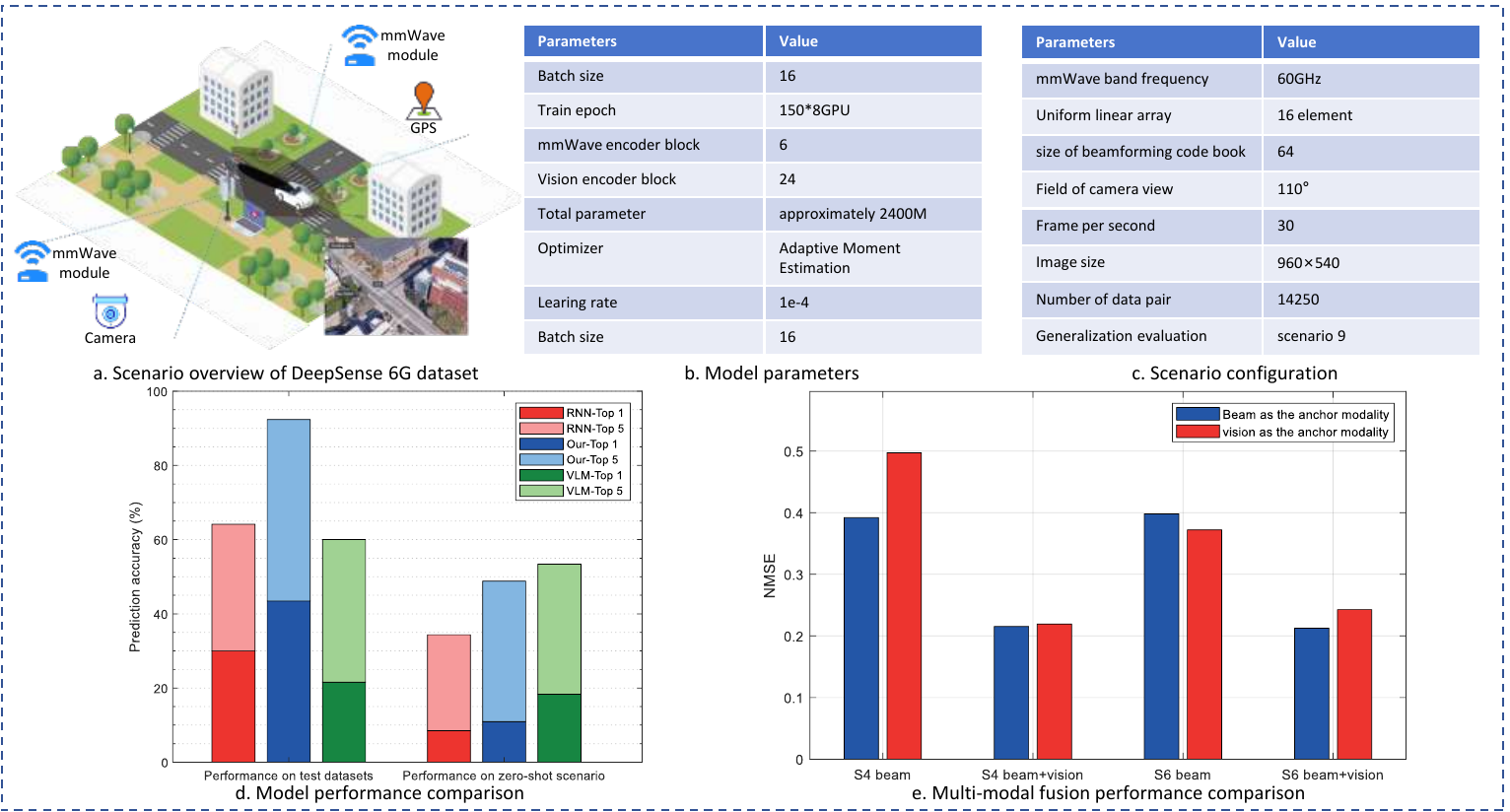}
  \caption{We conducted comparative experiments between our proposed paradigm and baseline models on the multi-modal beam prediction task, using real-world environmental datasets. (a)-(c) present the system environment and parameter configurations, while (d)-(e) demonstrate the experimental results.}
  \label{fig5}
\end{figure*}

Among these, the construction of the foundational model is of paramount importance. The potential challenge lies in how to provide customized foundational models for diversified network scenarios. To address this, we propose a two-stage training paradigm.

Firstly, the network selects data of matching modalities for processing based on the types of sensors and requirements in different scenarios, thereby constructing a dataset. Building on this, similar to related research paradigms~\cite{Clip,ImageBind}, the network selects a universal modality (such as signals, images, etc.) as the anchor point for multi-modal data alignment according to the specific scenario requirements, achieving alignment of data across different modalities. This is specifically accomplished by designing specialized encoders for different modalities, such as Vision Transformers (ViT), among others. Data from various modalities are transformed into corresponding embedded representations through these encoders. By employing contrastive learning methods, the data from other modalities can learn the latent associations with the anchor modality, resulting in a joint embedded representation space. Through the first stage of training, a pre-trained feature model that is generalizable within a specific scenario can be obtained.

To enable the model to be applied to various downstream tasks, the second stage involves constructing a model backbone and multiple task-specific heads to achieve task-specific adaptation. Specifically, a transformer-based backbone is built to efficiently scale up the model size and enhance its performance. Additionally, the introduction of task heads with different architectures equips the model with the capability to handle diverse wireless-native tasks, such as predictive tasks like beam prediction and classification tasks like human status detection. 

\section{Feasbility study}
\label{sec:5}

To validate the effectiveness of the proposed paradigm proposed in Section \ref{sec:5}, we constructed a large model with 2.4 billion parameters on a dataset collected from a real-world multi-modal ISAC system for verification.

\textbf{Environment Description:} We evaluated the proposed framework on the DeepSense 6G dataset~\cite{DeepSense}, which comprises extensive multi-modal sensing and communication data obtained from real-world measurements. For our experimental setup, we selected eight vehicle-to-infrastructure scenarios containing approximately 14,000 multi-modal data pairs for model training. Each scenario consists of two primary units: (1) a stationary unit serving as the base station, equipped with a millimeter-wave (mmWave) receiver, RGB camera, and GPS receiver; and (2) a mobile vehicle unit equipped with a mmWave transmitter and GPS receiver. The mmWave receiver features a 60GHz RF front-end with a 16-element uniform linear array and utilizes a combining beamforming codebook of 64 beams. The detailed scenario descriptions and system configurations are illustrated in Fig.\ref{fig5}.

\textbf{Task Description:} Multi-modal sensing-aided beam prediction is a promising task in mmWave communications, capable of enhancing the performance of beam management. To validate the scalability and generalization capability of the MLM, we defined a beam prediction task that utilizes a sequence of multi-modal sensing data from five time steps to predict the beams for the subsequent three time steps. To validate the effectiveness of wireless-native large-parameter models, we constructed a GPT-like model based on our proposed method, with specific parameter configurations illustrated in Fig.~\ref{fig5}. We use a Recurrent Neural Network (RNN) model with small-scale parameters and the Qwen-2.5-7B~\cite{Qwen2.5-VL} as large-scale MLLM baselines.

\textbf{Experimental result:} We selected beam, visual data as multi-modal inputs to verify the accuracy of beam prediction. The experimental results are shown in Fig.~\ref{fig5}d. Compared with small models, both WMLM and MLM demonstrate excellent performance and zero-shot capability. Moreover, due to training on a large amount of wireless network data, WMLM achieves the best performance on the test set, with generalization ability close to large-parameter MLM baseline.

To further evaluate the efficacy of different modality fusion methods, we employed beam signals and visual data as respective anchors for contrastive learning. Models were trained using datasets from four (S4) and six (S6) scenarios. This process aims to minimize the discrepancy between the model's predictions and the ground truth. As shown in Fig~\ref{fig5} e, representations aligned with either anchor modality outperform the baseline, and their joint representation yields the best results. Notably, using beam sequences as the anchor achieves a lower NMSE in most cases. This is expected, as wireless signals can serve as a network-native modality alignment paradigm, especially in the absence of visual data.

\section{Future directions and challenges}
\label{sec:6}
\textbf{Large-scale real world dataset:} The development of large-parameter models significantly enhances capabilities in accordance with scaling laws, while the integration of multi-modal data improves comprehensive understanding of the physical world and facilitates various communication and sensing tasks. However, unlike fundamental AI applications such as natural language processing and computer vision, the training of wireless-native foundation models faces a critical challenge: the scarcity of large-scale datasets, particularly those derived from real-world scenarios. Although numerous pioneering efforts have been made to construct datasets to advance AI for ISAC, there remains a notable absence of comprehensive, large-scale datasets encompassing diverse real-world scenarios with extensive data generated from sensors and communication modules. In future networking paradigms, emerging technologies such as digital twins hold promise for addressing this data scarcity issue.

\textbf{Explainable model:} Explainability is a critical issue in the field of AI. Understanding the reasons behind MLMs' predictions and decisions helps build a robust technical theoretical framework and promotes the practical application of models. In intelligent ISAC, especially in scenarios such as connected vehicles and industrial IoT, MLMs must be highly reliable and explainable to avoid unpredictable errors, such as incorrect sensing location judgments or unreasonable data predictions, which could endanger human safety or disrupt critical production activities. Although research in this area remains limited, it is crucial for the deployment of models in critical scenarios.

\textbf{Secure and private:} The data used for model training is fully exposed to potential attackers due to the open nature of wireless channels. Adversarial attacks, such as data poisoning, can compromise the robustness of MLMs, especially when the scale of training data is extremely large, making it significantly more challenging to inspect the data. Additionally, due to the sensitivity of time-series data, models may memorize specific details of the training data, posing risks of data leakage. Current wireless network architectures were not designed with AI security, particularly the security of MLMs. This necessitates the use of technologies such as differential privacy or federated learning to enhance the security of MLMs.

\section{Conclusion}
\label{sec:7}

This paper investigates WMLMs and elaborates on their key characteristics and design principles from the perspectives of multimodal data fusion and multi-task representation. Inspired by these fundamental principles, we propose a two-stage paradigm to demonstrate the training and deployment of WMLMs. Furthermore, we conduct feasibility study to validate the effectiveness of our approach, analyzing the advantages of WMLMs compared to baselines on real-world ISAC datasets. The results demonstrate WMLMs' performance in cross-scenario generalization and multimodal fusion capabilities, suggesting their potential as a pathway for future networks. Finally, we discuss promising research directions and remaining challenges in this emerging field.

\vfill


\begin{thebibliography}{1}
\bibliographystyle{IEEEtran}

\bibitem{LLM4Beam}
Y. Sheng, K. Huang, L. Liang, P. Liu, S. Jin and G. Ye Li, "Beam Prediction Based on Large Language Models," in IEEE Wireless Communications Letters, vol. 14, no. 5, pp. 1406-1410, May 2025.

\bibitem{LLM4CP}
B. Liu, X. Liu, S. Gao, X. Cheng and L. Yang, "LLM4CP: Adapting Large Language Models for Channel Prediction," in Journal of Communications and Information Networks, vol. 9, no. 2, pp. 113-125, June 2024.

\bibitem{CSI_BERT}
Z. Zhao, T. Chen, F. Meng, H. Li, X. Li and G. Zhu, "Finding the Missing Data: A BERT-Inspired Approach Against Package Loss in Wireless Sensing," IEEE INFOCOM 2024 - IEEE Conference on Computer Communications Workshops (INFOCOM WKSHPS), Vancouver, BC, Canada, 2024, pp. 1-6.

\bibitem{IMT-2030}
ITU-WP5D, “Framework and overall objectives of the future development of IMT for 2030 and beyond.”, 2023 [Online]. Available:\url{https://www.itu.int/dms_pubrec/itu-r/rec/m/R-REC-M.2160-0-202311-I!!PDF-E.pdf}.

\bibitem{multi-modal sensing and comm}
Cheng, Xiang, et al. “Intelligent Multi-Modal Sensing-Communication Integration: Synesthesia of Machines.” IEEE Communications Surveys and Tutorials/IEEE Communications Surveys and Tutorials, vol. 26, no. 1, 1 Jan. 2024, pp. 258–301.

\bibitem{timer}
Y. Liu, H. Zhang, C. Li, X. Huang, J. Wang, and M. Long, ``Timer: Generative Pre-trained Transformers Are Large Time Series Models,'' in Proceedings of the 41st International Conference on Machine Learning, 2024, pp. 1--31.

\bibitem{GenAI4PHY}
N. V. Huynh et al., “Generative AI for Physical Layer Communications: A Survey,” IEEE Transactions on Cognitive Communications and Networking, vol. 10, no. 3, pp. 706–728, Jun. 2024.

\bibitem{multi-modal survey}
T. Baltrusaitis, C. Ahuja, and L.-P. Morency, “multi-modal Machine Learning: A Survey and Taxonomy,” IEEE Transactions on Pattern Analysis and Machine Intelligence, vol. 41, no. 2, pp. 423–443, Feb. 2019.

\bibitem{ImageBind}
R. Girdhar, A. El-Nouby, Z. Liu, M. Singh, K. V. Alwala, A. Joulin, I. Misra*, "ImageBind: One Embedding Space To Bind Them All," \textit{To appear in IEEE/CVF Conference on Computer Vision and Pattern Recognition (CVPR)}, 2023.

\bibitem{Clip}
A. Radford, J. W. Kim, C. Hallacy, A. Ramesh, G. Goh, S. Agarwal, G. Sastry, A. Askell, P. Mishkin, J. Clark et al., "Learning Transferable Visual Models From Natural Language Supervision," \textit{In Proceedings of the 38th International Conference on Machine Learning (ICML)}, pp. 8748–8763, 2021.

\bibitem{LLM4WM}
X. Liu, S. Gao, B. Liu, X. Cheng, and L. Yang, ``LLM4WM: Adapting LLM for Wireless Multi-Tasking,'' \textit{arXiv preprint arXiv:2501.12983}, 2025.

\bibitem{NetLLM}
D. Wu, X. Wang, Y. Qiao, Z. Wang, J. Jiang, S. Cui, and F. Wang, ``NetLLM: Adapting Large Language Models for Networking,'' in \textit{Proceedings of the ACM SIGCOMM 2024 Conference}, 2024.

\bibitem{Agent}
X. Cao, G. Nan, H. Guo, H. Mu, L. Wang, Y. Lin, Q. Zhou, J. Li, B. Qin, Q. Cui, X. Tao, H. Fang, H. Du, and T. Q. S. Quek, ``Exploring LLM-Based Multi-Agent Situation Awareness for Zero-Trust Space-Air-Ground Integrated Network,'' \textit{IEEE Journal on Selected Areas in Communications}, vol. 43, no. 6, pp. 2230-2247, June 2025.

\bibitem{DeepSense}
A. Alkhateeb, G. Charan, T. Osman, A. Hredzak, J. Morais, U. Demirhan, and N. Srinivas, ``DeepSense 6G: A Large-Scale Real-World Multi-Modal Sensing and Communication Dataset,'' \textit{IEEE Communications Magazine}, vol. 61, no. 9, pp. 122-128, September 2023.

\bibitem{Qwen2.5-VL}
S. Bai et al., ``Qwen2.5-VL Technical Report,'' \textit{arXiv preprint arXiv:2502.13923}, 2025.

\end{thebibliography}
\end{document}